\documentclass[12pt,preprint]{aastex}

\usepackage{graphicx}
\usepackage[space]{grffile}
\usepackage{latexsym}
\usepackage{textcomp}
\usepackage{longtable}
\usepackage{multirow,booktabs}
\usepackage{amsfonts,amsmath,amssymb}
\usepackage{bm}
\usepackage[utf8]{inputenc}
\usepackage[english]{babel}

\newcommand{\pflux}{\rm{phot~cm$^{-2}$~s$^{-1}$}}

\newcommand{\be}{\begin{equation}}
\newcommand{\ee}{\end{equation}}
\newcommand{\ba}{\begin{eqnarray}}
\newcommand{\ea}{\end{eqnarray}}

\newcommand{\simgt}{\lower 2pt \hbox{$\, \buildrel {\scriptstyle >}\over {\scriptstyle\sim}\,$}}
\newcommand{\simlt}{\lower 2pt \hbox{$\, \buildrel {\scriptstyle <}\over {\scriptstyle\sim}\,$}}
\newcommand{\ls}{\lower 2pt \hbox{$\;\scriptscriptstyle \buildrel<\over\sim\;$}}
\newcommand{\gs}{\lower 2pt \hbox{$\;\scriptscriptstyle \buildrel>\over\sim\;$}}

\newcommand{\fermi}{\emph{Fermi}}
\newcommand{\iras}{\emph{IRAS}}
\newcommand{\degree}{$^\circ$}
\newcommand{\degrr}{$^{\circ}\!\!.$}

\usepackage{csquotes}

\slugcomment{Submitted to ApJ}

\begin{document}

\title{Constraining Gamma-Ray Emission from Luminous Infrared Galaxies with Fermi-LAT; Tentative Detection of Arp 220}

\author{Rhiannon D. Griffin\altaffilmark{1}, Xinyu Dai\altaffilmark{1}, Todd A. Thompson\altaffilmark{2}} 

\altaffiltext{1}{Homer L. Dodge Department of Physics and Astronomy,
University of Oklahoma, Norman, OK, 73019;
Rhiannon.D.Griffin-1@ou.edu, xdai@ou.edu}
\altaffiltext{2}{Department of Astronomy and Center for Cosmology and Astro-Particle Physics, 
The Ohio State University, Columbus, Ohio 43210; 
thompson@astronomy.ohio-state.edu}

\begin{abstract}
Star-forming galaxies produce gamma-rays primarily via pion production, resulting from inelastic collisions between cosmic ray protons and the interstellar medium (ISM). The dense ISM and high star formation rates of luminous and ultra-luminous infrared galaxies (LIRGs and ULIRGs) imply that they should be strong gamma-ray emitters, but so far only two LIRGs have been detected. Theoretical models for their emission depend on the unknown fraction of cosmic ray protons that escape these galaxies before interacting. We analyze \fermi-LAT data for 82 of the brightest \iras\ LIRGs and ULIRGs. We examine each system individually and carry out a stacking analysis to constrain their gamma-ray fluxes. We report the detection of the nearest ULIRG Arp 220 ($\sim4.6\sigma$).
We observe a gamma-ray flux (0.8--100 GeV) of $2.4 \times 10^{-10}$\,phot cm$^{-2}$~s$^{-1}$ with photon index of 2.23 ($8.2 \times 10^{41}$\,ergs s$^{-1}$ at 77 Mpc) We also derive upper limits for the stacked LIRGs and ULIRGs. The gamma-ray luminosity of Arp~220 and the stacked upper limits agree with calorimetric predictions for dense star-forming galaxies. With the detection of Arp 220, we extend the gamma-ray--IR luminosity correlation to the high luminosity regime with $\log{L_{0.1-100 \textrm{GeV}}} = 1.25\times\log{L_{8-1000 \mu\textrm{m}}} + 26.7$ as well as the gamma-ray--radio continuum luminosity correlation with $log{L_{0.1-100 \textrm{GeV}}} = 1.22\times\log{L_{1.4 \textrm{GHz}}} + 13.3$. The current survey of \fermi-LAT is on the verge of detecting more LIRGs/ULIRGs in the local universe, and we expect even more detections with deeper \fermi-LAT observations or the next generation of gamma-ray detectors. 
\end{abstract}


\section{Introduction}

Gamma-ray emission provides a sensitive probe of the cosmic ray content of star-forming galaxies.
When cosmic ray protons collide with the dense ISM of star-forming galaxies, they produce secondary electron-positron pairs, neutrinos, and gamma-rays through pion production on a characteristic timescale
$t_{\rm pp}\simeq7\times10^7\,{\rm yr}\,n^{-1},$
where $n$ is the gas density of the ISM in units of ${\rm cm}^{-3}$ (e.g., \citealt{Schlickeiser_2002}). For average ISM densities larger than $\sim10-100$\,cm$^{-3}$, $t_{\rm pp}$ may be sufficiently short that one expects most of the cosmic rays to interact with the ISM before escaping the host galaxy through diffusion, or via advection in a large-scale galactic wind \citep{Loeb_Waxman}. Assuming that cosmic rays are predominantly accelerated in supernovae, if the cosmic ray escape time is much longer than $t_{\rm pp}$, then one expects a one-to-one linear relation between the gamma-ray luminosity and the star formation rate, as measured by the far-infrared luminosity or GHz radio continuum (RC) luminosity \citep{Thompson_2007,Lacki_2010,Lacki_2011}. 

The diffuse gamma-ray emission from star-forming galaxies and its contribution to the extragalactic diffuse gamma-ray background has been predicted by a number of authors \citep[e.g.,][]{Paglione_1996,Blom_1999,Torres_2004,Cillis_2005,Thompson_2007,Pavlidou_2001}. 
Observational breakthroughs occurred with the detections of nearby starburst galaxies M82 \citep[VERITAS,][]{Acciari_2009} and NGC~253 \citep[H.E.S.S.,][]{Acero_2009} by Cherenkov telescopes and \fermi\ \citep{Abdo_2010}.  
\citet{Ackermann_2012} further summarized the \fermi\ detections of diffuse gamma-ray emission from local group galaxies, including the Milky Way, M31, the LMC and SMC, and from nearby star-forming galaxies NGC~4945 and NGC~1068. \citet{tang} observed a $\sim5.5\sigma$ detection from the luminous infrared galaxy NGC~2146. \citet{Ackermann_2012} also established empirical correlations between the diffuse gamma-ray emission and both the FIR or radio emission of star-forming galaxies, where the latter two continua are also tightly correlated \citep[e.g.,][]{Vanderkruit_1971,Yun_2001,Sargent_2010,Bourne_2011}.

In this paper, we focus on \fermi\ observations of luminous (LIRGs; $10^{11}<L_{\rm IR}(8-1000\mu{\rm m})/L_{\odot}<10^{12}$) and ultra-luminous IR galaxies  (ULIRGs (LIRGs; $L_{\rm IR}(8-1000\mu{\rm m})/L_{\odot}>10^{12}$), probing the highest luminosity regime.  We assume a flat $\Lambda$CDM cosmology of $H_0 = 70~{\rm km~s^{-1} Mpc}^{-1}$, $\Omega_m = 0.3$, and $\Omega_\Lambda=0.7$.

\section{Analysis} 
For our galaxy sample, we use infrared bright ($L_{\rm IR}(8-1000 \mu m) > 10^{11} L_{\odot}$) galaxies from the \textit{Infrared Astronomical Satellite} (\iras) Revised Bright Galaxy Sample (RBGS, \citealt{Sanders_2003}). 
The RBGS contains all 629 extragalactic objects brighter than 5.24 Jy at 60 $\mu$m surveyed by the \iras. These are the brightest extragalactic 60 $\mu$m sources \citep{Sanders_2003} and are ideal candidates for studying gamma-ray emission in star-forming galaxies.
Most of the galaxies have low infrared luminosities $L_{\rm IR}(8-1000 \mu m) < 10^{11} L_{\odot}$. 
Further excluding targets close to the Galactic plane ($\lvert b \rvert < 20$\degree), where the Galactic gamma-ray background is high, our sample contains 135 galaxies with a range of redshifts from 0.0030 to 0.082, with a median redshift 0.022. 
In this sample, there are 123 LIRGs and 12 ULIRGs.
As discussed later in this section, we use 82 of the total 135 galaxies by further excluding targets with high backgrounds. 

Using 399 weeks ($\sim7.7$ years) of \fermi-LAT data, we obtain photon flux upper limits (ULs) from individual and stacked photon count maps of galaxy positions.
We follow a similar method as described in \citet{Griffin_2014}, where we study \fermi-LAT count map stacks of galaxy clusters to obtain the lowest ULs to date. 
Here we summarize the method and discuss any changes and updates that are tailored to stack galaxies. For more details, see \citet{Griffin_2014}.
In general, galaxies have much smaller angular sizes than nearby galaxy clusters, and
considering the angular resolution of \fermi-LAT, most galaxies are unresolved in the gamma-ray regime.
Therefore, we choose to use a fixed angular radius to stack the data rather than a fixed physical radius as in \citet{Griffin_2014}.  The corresponding background regions are in general smaller, and the number of contaminating background sources lower. 
We use PASS8 LAT source photons and determine good time intervals using the recommended expression \texttt{((DATA\_QUAL==1) \&\& (LAT\_CONFIG==1))}. 
We use the recommended zenith angle cut of 90\degree\ to exclude gamma-rays from Earth's atmospheric limb.

For each count map, we exclude weeks with high background flaring activity, where the photon flux (phot~cm$^{-2}$~s$^{-1}$) is greater than 2$\sigma$ above the mean flux. This removes the contribution to the background from variable gamma-ray sources \citep{Griffin_2014}.
We use 7 logarithmically spaced energy bins across the bandpass $0.8-100$ GeV.
We use a pixel size of 0\degrr4
and count maps of 4\degree\ radii to estimate the background contribution. 
We tested other pixel radii; however they were either much lower than the resolution of the \fermi-LAT, or were too large, 
which results in a larger background, increasing the number of contaminated targets.
The 4\degree\ count map radius was found to be optimal to encompass source and background regions. 
The lower energy bin PSFs ($\sim1$ GeV) are larger than the 0\degrr4
source radius, so we calculate an energy-dependent aperture flux correction to apply to flux limits calculated in all energy bands.

\begin{figure}
\begin{center}
\includegraphics[width=0.938\columnwidth]{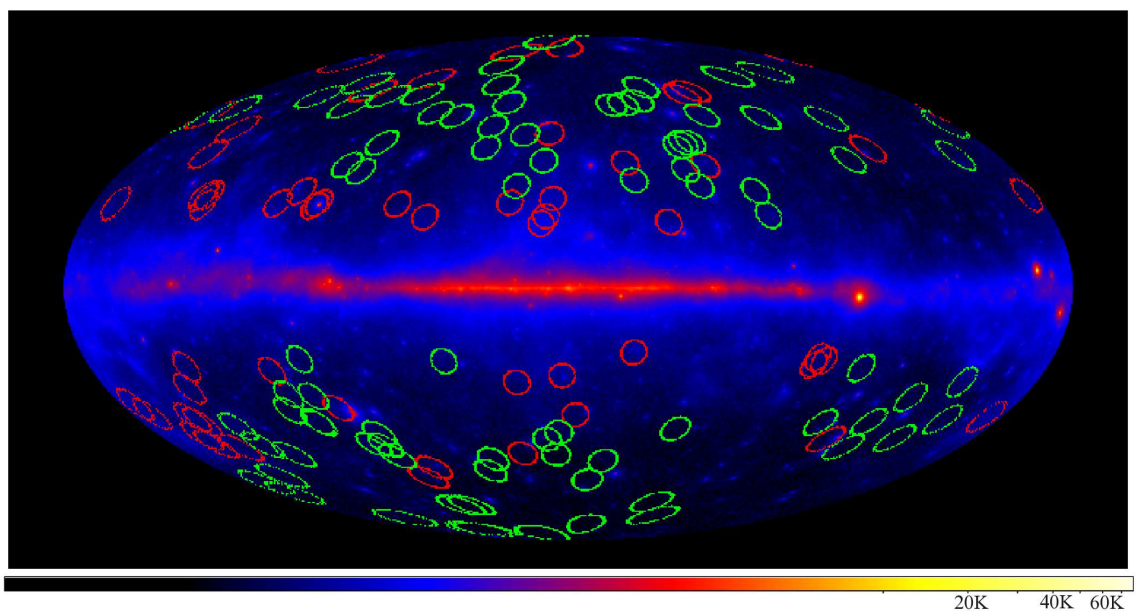}
\caption{\label{fig:allsky}
All-sky gamma-ray photon count map, 135 ellipses indicate positions of \iras\ galaxies used in this analysis and their 4\degree\ extents on the sky. The 82 green ellipses shown include Arp 220 and the 81 galaxies used in the final stacked image. Red ellipses are rejected sources due to background contamination.%
}
\end{center}
\end{figure}

We first co-add the exposure and count maps in the time domain so that we have one count map and one exposure map for each galaxy position.
We flatten each count map using the average exposure measured from the center so that the effective exposure time is uniform across each image.
To minimize background contamination from known point sources, we mask bright sources from the 3FGL catalog \citep{3fgl}.  We exclude any photons within 0\degrr5 of the source and populate this region with randomly placed photons using the average local background measured in the annulus with inner and outer radii 0\degrr7 to 0\degrr9, respectively.
We visually rejected any count maps with residual contamination from poorly masked 3FGL sources. 
In general, these contaminating sources are nearby blazars or pulsars.
Among the rejected cases are the previously detected LIRGs NGC 2146 and NGC 1068 \citep{tang,Ackermann_2012}.
In our final sample, we have 82 galaxies where 7 are ULIRGs, including Arp 220 and Mrk 273. The other 75 are LIRGs.  The positions of these galaxies and their 4\degree\ extent on the sky are shown in green in Figure \ref{fig:allsky}, while the rejected galaxies are shown in red. 

\section{Results and Discussion}
\label{sec:results}

\begin{figure}
\begin{center}
\includegraphics[width=0.98\columnwidth]{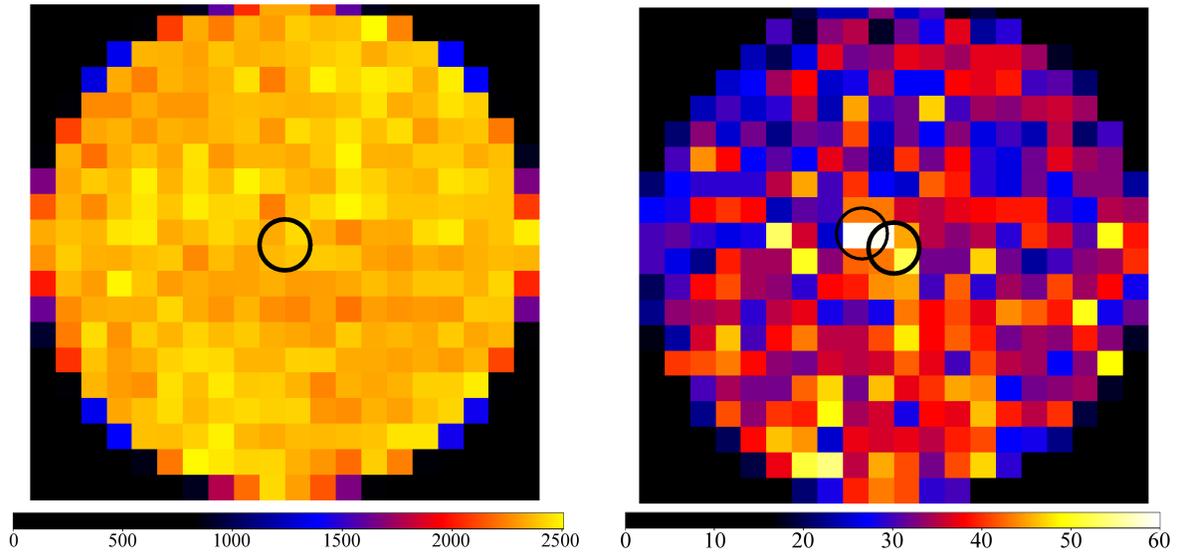}
\caption{\label{fig:cmaps}
Left: Final stacked photon count map of 81 \iras\ galaxies. We exclude the detection of Arp 220 from this stack. Right: Photon count map of Arp 220 with $3.78\sigma$ source detection. The thinner black circle indicates the position of CRATES J$153246+234400$, the likely source of contamination. A binned likelihood analysis determined a TS value of 21.3 ($\sim4.6\sigma$) for Arp 220. Both count maps include a black circle indicating the 0\degrr4 source region. %
}
\end{center}
\end{figure}

\begin{figure}
\center
\includegraphics[width=1.0\textwidth]{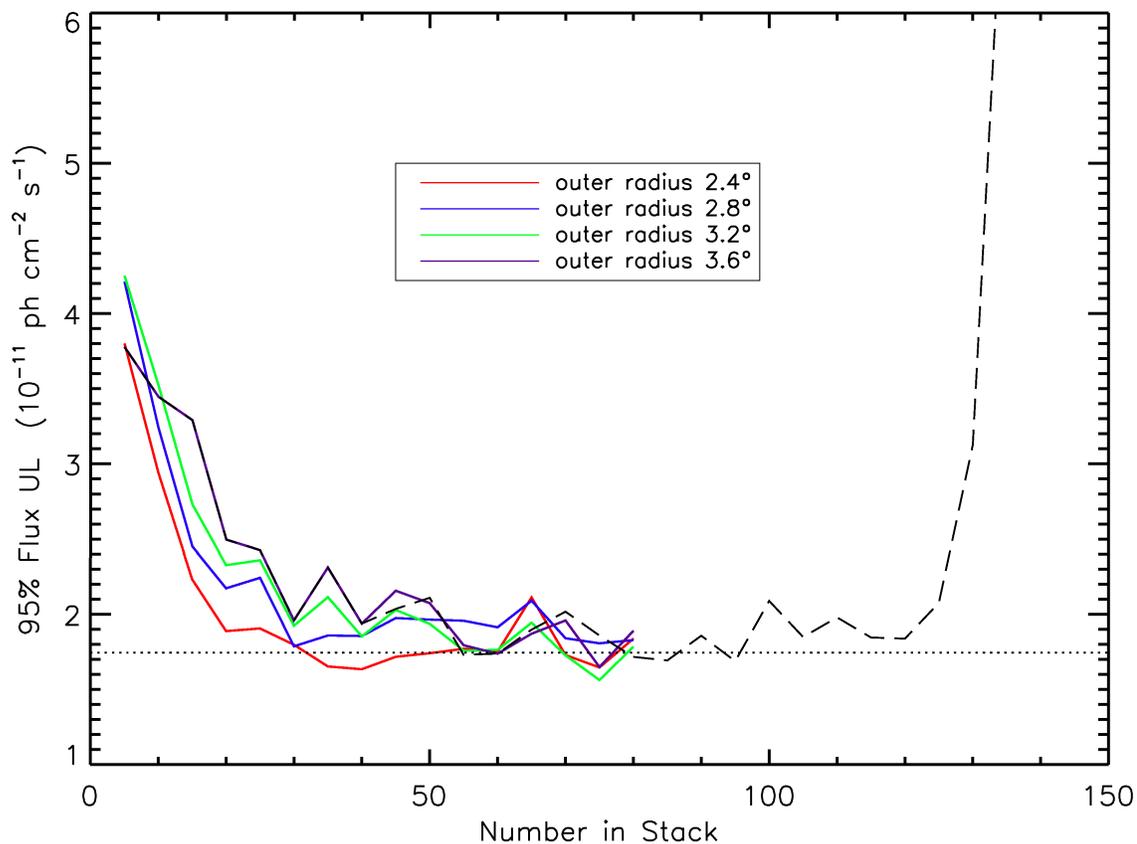}
\caption{\label{fig:fluxvsstack}
Photon flux upper limits versus stack size, sorted by background variance. The solid lines represent the four diffent outer radii used for the 81 count maps in the final stack (we exclude the detection of Arp 220). As the stack size increases, the flux ULs decrease and flatten as expected. The dashed line represents the stacking analysis of the 135 count maps,  including those with background contaminations from 3FGL sources, namely blazars and pulsars. The horizontal dotted line is the median flux of the final 81 count maps stack: $1.74 \times 10^{-11}$ phot cm$^{-2}$ s$^{-1}$.%
}
\end{figure}

Figure \ref{fig:cmaps} (left) shows the final stack of 81 photon count maps with a black central circle indicating the 0\degrr4 source region.  We detect no excess gamma-ray source emission above the background and find an aperture corrected 95\% confidence upper limit of $1.74 \times 10^{-11}$ \pflux\ per galaxy in the $0.8-100$ GeV energy band (excluding Arp 220).
ULs are obtained using a fixed inner background radius of 1\degrr2 and outer background radii of 2\degrr4, 2\degrr8, 3\degrr2, and 3\degrr6, respectively.
The UL of $1.74 \times 10^{-11}$ \pflux\ per galaxy is the median UL using the four background regions and is represented by the horizontal line in Figure \ref{fig:fluxvsstack}. 
We find that the choice of outer background radius has little impact on the UL, as shown in Figure \ref{fig:fluxvsstack}. 
We stack the count maps in order of background variance and as expected, as we add more photon count maps the flux ULs drop significantly and level out after $\sim50$. 
Additionally, Figure \ref{fig:fluxvsstack} shows the ULs obtained from the stack of 135 count maps, including the visually rejected point source contamination cases. This shows that we do not see improvement from including these count maps and that the choice to exclude these cases is valid. The sharp uptick at the right end of Figure~\ref{fig:fluxvsstack} is from the inclusion of 2 count maps that have the pulsar PSR J1836+5925 (e.g., \citealt{3fgl}), providing strong contamination to the background. We also study the LIRG and ULIRG stacks separately. 
The LIRG photon flux UL we obtain is $1.73 \times 10^{-11}$ \pflux\ per galaxy and with a mean redshift of 0.023 for the 75 galaxies, and we obtain a corresponding luminosity limit of $1.30 \times 10^{41}$ ergs s$^{-1}$ in the $0.8-100$ GeV energy band. 
The UL we obtain from stacking the 6 ULIRGs, excluding Arp 220, is  $5.44 \times 10^{-11}$ \pflux\ per galaxy. With a mean redshift of 0.052, we obtain a corresponding luminosity limit of $2.19 \times 10^{42}$ ergs s$^{-1}$ in the $0.8-100$ GeV energy band. 

In addition to examining stacks of galaxy count maps, we study each galaxy individually.
In all but one case, we do not detect any source significantly above the background. The one exception is Arp 220, the closest ULIRG to our galaxy, where we detect gamma-ray emission at $3.78\sigma$ significance above the local background in the $0.8-100$ GeV energy band. 
Excluding Arp 220, we find that the individual flux ULs for each galaxy in the entire sample range from $8.2 \times 10^{-11}$ to $3.0 \times 10^{-10}$ \pflux. 
Of the 81 galaxies, we measure a median individual galaxy UL of $1.4 \times 10^{-10}$ \pflux\ and mean redshift of 0.026. We obtain a corresponding luminosity limit of $1.3 \times 10^{42}$ ergs s$^{-1}$ in the $0.8-100$ GeV energy band. 

Arp 220 is a two galaxy merging system with extreme conditions, and has been studied extensively across multiple wavelengths \citep[e.g.,][]{Smith_1998,Heckman_1996,Dunne_2001,Rangwala_2011,Lacki_2011}. 
The redshift of Arp~220 is $z = 0.018$, implying a distance of 77 Mpc. 
This ULIRG has previously not been detected in the gamma-ray regime and one of the galaxy nuclei might house a hidden AGN \citep[e.g.,][]{Iwasawa_2001,Rangwala_2011}.
We detect a gamma-ray signal at the location of Arp~220 that is $3.78\sigma$ above the mean background level (Figure~\ref{fig:cmaps}, right), where we sample the mean background and its standard deviation by randomly drawing 500 regions of the same size as the source region in the background area.
The off-center peak emission seen in Figure~\ref{fig:cmaps} implies some contamination from a nearby source. This emission is not from a source in the 3FGL catalog \citep{3fgl}.
In their analysis of Arp 220, \citet{tang} suggest CRATES J$153246+234400$, a flat-spectrum radio quasar (FSRQ, \citealt{Healey_2007}) as the likely candidate of the contamination. The position of this source is located 0.55\degree\ from Arp 220 and aligns with the peak emission, as indicated in Figure~\ref{fig:cmaps} with the thinner black circle. We examine smaller energy bins to determine if the detection of Arp 220 is significant despite the contamination. We use the energy dependent mean PSFs of each sub energy bin as a source radius. We mask the overlapping area from the contaminating source and fill in that mask with a proportional count of photons from the unaffected area of the Arp 220 signal. In so doing, we find that the first energy bin with energies [0.8, 1.6] GeV is contaminated but the remaining energy bands together [1.6, 100] GeV provide a $3.03\sigma$ signal above the background. Thus, it seems there is a significant signal from Arp 220 and we analyze the system more closely. 

We perform a binned likelihood analysis of Arp 220 using the recommended method from the Fermi Science Support Center\footnote{FSSC, fermi.gsfc.nasa.gov/ssc/}. We use photons with energies in the bandpass 0.8--100 GeV, using an ROI of radius 10\degree\ with Arp 220 at the center and we model the emission from  all 3FGL sources within 15\degree\ as point sources. We include two additional point sources, for Arp 220 and CRATES J$153246+234400$, modelled with power-law spectrums: $dN/dE = N_{\circ}(E/E_{\circ})^{-\Gamma_{ph}}$, where $\Gamma_{ph}$ is the photon index. We model the Galactic and extragalactic backgrounds using the most up-to-date versions: gll\_psc\_v16.fit and iso\_P8R2\_SOURCE\_V6\_v06.txt.
For CRATES J$153246+234400$, we measure a TS value, photon index, and photon flux of 20.3 (4.5$\sigma$), $3.472 \pm 1.23$ and $2.66 \pm 0.78 \times 10^{-10}$, respectively.
For Arp 220, we measure a TS value of 21.3 (4.6$\sigma$) with a photon index of $2.23 \pm 0.46$ and photon flux $2.43 \pm 0.90 \times 10^{-10}$ \pflux\ in the energy band [0.8, 100] GeV. Using the measured photon index, we find the luminosity of Arp 220 to be $8.22 \pm 3.0  \times 10^{41}$ ergs s$^{-1}$. Independent of our analysis, \citet{Peng_2016} also report a gamma-ray detection of Arp 220 ($\sim6.3\sigma, \Gamma_{ph} = 2.35 \pm 0.16$) in the energy band [0.2, 100] GeV, reported concurrently with this work. Their gamma-ray luminosity of $L_{0.1-100 \textrm{GeV}} = 1.78 \pm 0.30 \times 10^{42}$ ergs s$^{-1}$ is consistent with our value of $L_{0.1-100 \textrm{GeV}} = 1.57 \pm 0.58 \times 10^{42}$ ergs s$^{-1}$.

The detection of Arp 220 is largely in agreement with previous theoretical models, including explicitly proton calorimetric estimates \citep{Thompson_2007}, and more detailed treatments \citep[e.g.,][]{Torres_2004,Lacki_2010,Lacki_2011,Lacki_2012,Yoast_Hull_2015}.
The implications of the detection of Arp 220 in the gamma-ray regime, with a luminosity compatible with the calorimetric limit, have been described extensively by \citet{Lacki_2011}. Here, we briefly summarize. The high gamma-ray luminosity implies (1) a low equilibrium energy density for cosmic rays with respect to the energy density required for hydrostatic equilibrium, (2) secondary electron/positron pairs from pion production likely dominate production of the observed GHz radio continuum 
(\citealt{Torres_2004}, \citealt{Rengarajan_2005}, see eq.~19 of \citealt{Lacki_2011}), (3) relativistic bremsstrahlung and ionization losses flatten the continuum synchrotron spectrum (see \citealt{Thompson_2006}), potentially providing evidence for the \enquote{high gas surface density} conspiracy for the FIR-radio correlation described in \citet{Lacki_2010}, and finally (4) star-forming galaxies contribute to the diffuse gamma-ray and neutrino backgrounds 
 \citep[e.g.,][]{Pavlidou_2001,Loeb_Waxman,Thompson_2007,Lacki_2014,Murase_2013}. Gamma-rays with $>$\,TeV energies are expected to be attenuated 
 \citep{Torres_2004,Lacki_2012,Yoast_Hull_2015}, producing high-energy electron/positron pairs that may contribute to the observed diffuse X-ray emission via synchrotron radiation \citep{Lacki_2012}.

\begin{figure*}
\begin{center}
\includegraphics[width=1\columnwidth]{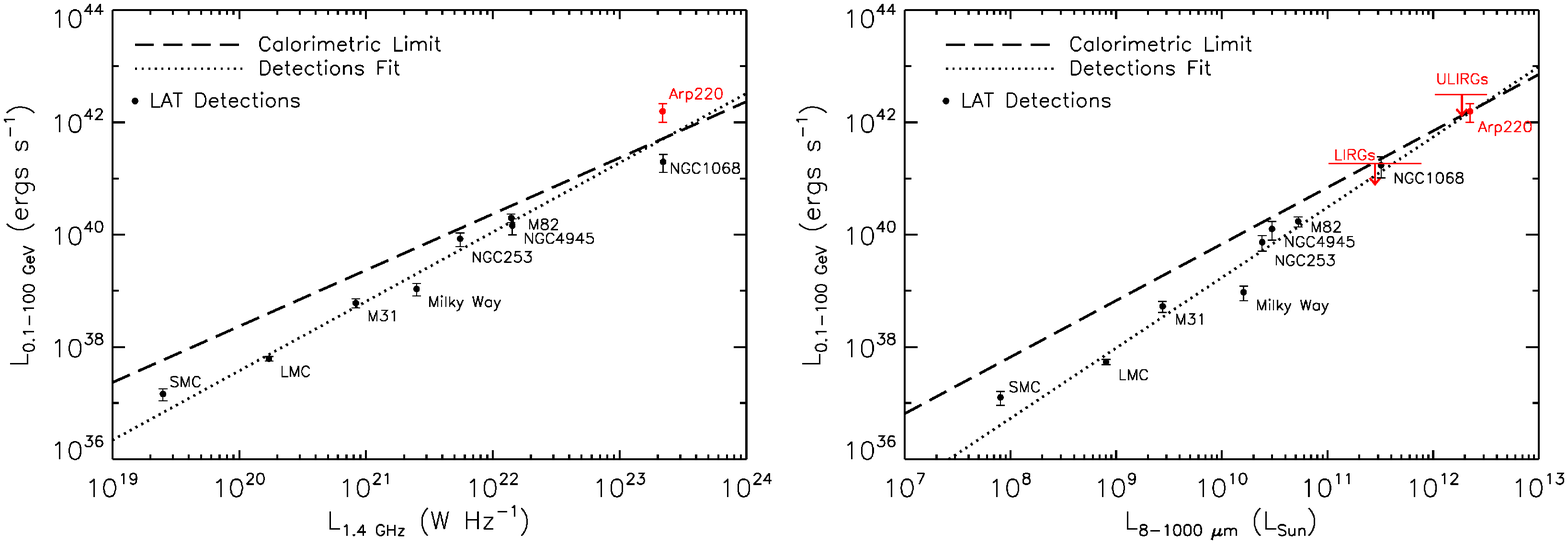}
\caption{\label{fig:lumin}
Gamma-ray luminosity versus radio continuum luminosity (left) and infrared luminosity (right), based off Figure~4 from \citet{Ackermann_2012}. The dashed line is the expected gamma-ray luminosity in the calorimetric limit \citep{Thompson_2007,Lacki_2011,Ackermann_2012}. The points represent \emph{Fermi}-LAT detections, where the most luminous is our detection of Arp 220, the rest are from \citet{Ackermann_2012}. The dotted line is the best fit to the detections using a $\chi^2$ best fit. Our Arp 220 detection agrees well with the expected signal based off of previous \fermi-LAT detections. The red arrows indicates our upper limits determined by the stack of 75 LIRGs and the stack of 6 ULIRGs (excluding the detection of Arp 220). The red horizontal stripes indicate the infrared luminosity bins used in the stacks.%
}
\end{center}
\end{figure*}

\citet{Ackermann_2012} examined correlations in star-forming galaxies between the gamma-ray luminosity and other tracers of star formation histories. Total IR luminosity (8--1000 $\mu$m) is one such tracer, as the ultraviolet light from massive stars is absorbed by dust and re-radiated as infrared (e.g., \citealt{Kennicutt_1998}). Another tracer is the RC luminosity that originates from CR electrons and positrons producing synchrotron radiation. 
With our gamma-ray flux measurements of Arp~220, we can extend the correlation to the high luminosity end.
Figure \ref{fig:lumin} compares gamma-ray luminosities to these other tracers of star formation history, RC luminosity on the left and total IR luminosity on the right. In the calorimetric limit a power-law relationship is expected in either case, represented by the dashed lines (see \citealt{Ackermann_2012} for details). We include \fermi-LAT detections from the local group \citep{Abdo_2010m31m33,Abdo_2010smc,Abdo_2010lmc}, the four detections by \citet{Ackermann_2012}, and our detection of Arp 220. Additionally, we include our upper limits from the LIRG and ULIRG stacks.
For this plot, we convert our luminosities found in the [0.8 -- 100] GeV bandpass to the luminosity band of [0.1 -- 100] GeV used by \citet{Ackermann_2012}, assuming a power-law spectral shape,
$dN/dE = N_{\circ}(E/E_{\circ})^{-\Gamma_{ph}}$
and $\Gamma_{ph} = 2$ (e.g., \citealt{Lacki_2011}).  
For Arp 220, we use the measured photon index of 2.23.
Furthermore, we convert the \citet{Ackermann_2012} luminosities from their adopted Hubble constant of 75 km s$^{-1}$ Mpc$^{-1}$ to our adopted value of 70 km s$^{-1}$ Mpc$^{-1}$. The measurements from this paper are shown in red in Figure \ref{fig:lumin}. We fit the detections using $\chi^2$ minimization assuming a simple power-law relation. The gamma-ray--IR luminosity correlation is described by the best fit line: 
\begin{equation}
\log{L_{0.1-100 \textrm{GeV}}} = (1.25 \pm 0.03)\times \log{L_{8-1000 \mu\textrm{m}}} + (26.7 \pm 0.29),
\end{equation}
as well as the gamma-ray--RC luminosity correlation with
\begin{equation}
\log{L_{0.1-100 \textrm{GeV}}} = (1.22 \pm 0.03)\times \log{L_{1.4 \textrm{GHz}}} + (13.3 \pm 0.58).
\end{equation}

Our Arp 220 detection lies right on the line for the gamma-ray-FIR fit and above the gamma-ray-RC fit but within the uncertainties. 

As \fermi-LAT collects more data, we expect to see more detections of nearby ULIRGs and LIRGs. Our ULs from the LIRG and ULIRG stacks are approaching the power-law fit from the detections. If we treat this fit as a detection threshold, we would only need to improve our LIRG UL by a factor of two and our ULIRG UL by a factor of three to cross this threshold. 

In this Letter, we have applied the method of \citet{Griffin_2014} of making and stacking \fermi-LAT count maps to study LIRGs and ULIRGs. We present upper limits as well as a $4.6\sigma$ detection of the previously undetected Arp 220, the closest ULIRG. We compare these to similar studies, namely those presented in \citet{Ackermann_2012}. Our results place further constraints on expected gamma-ray emission from these star-forming galaxies in the more energetic regime. We show that our upper limits are close to the current detection limit of \fermi-LAT and expect to see more detections in the next few years.

\acknowledgements
This research has made use of the publicly available \fermi-LAT data. 
We thank the anonymous referee for helpful comments. We acknowledge the financial support from the NASA ADAP programs NNX11AD09G, NNX15AF04G, and NSF grant AST$-$1413056. 
TAT is supported by NSF Grant \#1516967.  TAT thanks Brian Lacki, Eliot Quataert, and Eli Waxman for discussions and collaboration.

\end{document}